\documentclass[aps,prl,reprint,twocolumn]{revtex4-2}
\usepackage{amsmath,amssymb,amsthm,hyperref,graphicx}
\hypersetup{hidelinks}
\usepackage[capitalise]{cleveref}
\hypersetup{colorlinks=true, linkcolor=blue, citecolor=red, urlcolor=blue}

\begin{document}
\title{Enhanced magnon spin current using the bosonic Klein paradox}
\author{J. S. Harms$^{1}$}
\email{j.s.harms@uu.nl}

\author{H. Y. Yuan$^{1}$}
\email{huaiyangyuan@gmail.com}

\author{Rembert A. Duine$^{1,2}$}
\affiliation{$^{1}$Institute for Theoretical Physics, Utrecht University, 3584CC Utrecht, The Netherlands}
\affiliation{$^{2}$Department of Applied Physics, Eindhoven University of Technology, P.O. Box 513, 5600 MB Eindhoven, The Netherlands}
\date{\today}

\begin{abstract}
Efficient manipulation of magnons for information processing is a central topic in spintronics and magnonics. An outstanding challenge for long-distance spin transport with minimal dissipation is to overcome the relaxation of magnons and to amplify the spin current they carry.
Here, we propose to amplify magnon currents based on the realization of the bosonic Klein paradox in magnetic nanostructures. This paradox involves the magnon's antiparticle, the antimagnon, of which the existence is usually precluded by magnetic instabilities as it is an excitation at negative energy.
We show that, by appropriately tuning the effective dissipation through spin-orbit torques, both magnons and antimagnons are dynamically stabilized. As a result, we find that the reflection coefficient of incident magnons at an interface between two coupled magnets can become larger than one, thereby amplifying the reflected magnon current.
Our findings can lead to magnon amplifier devices for spintronic applications. Furthermore, our findings yield a solid-state platform to study the relativistic behavior of bosonic particles, which is an outstanding challenge with fundamental particles.
\end{abstract}

\maketitle
{\it Introduction.---} Spin waves, and their quanta called magnons, are collective excitations that occur in ordered magnets.
The emerging field of magnonics utilizes magnons for information processing \cite{Chumak2015}. As information carriers, magnons have the advantages of low power-consumption and efficient parallel data processing, as they do not give rise to Joule heating.
Furthermore, they are useful for both classical information processing, which includes logic gates \cite{Kostylev2005,Ganzhorn2016}, transistors \cite{Chumak2014,Wu2018,Corn2018,Joel2018} and diodes \cite{Lan2015}, and for quantum science and technologies, including single-magnon states, squeezed states and entanglement with other quantum platforms \cite{Yuan2020,Kamra2020,Lach2021,Yuanreview}.
A hurdle towards realizing magnon-based technology is the dissipation of magnons which results from interactions of magnons with their environment, such as conduction electrons, phonons and impurities.
These interactions dissipate the amplitude and coherence of magnon currents and are detrimental for  efficient application of magnons in nanoscale spintronic devices.
Therefore, a central challenge in magnonics is to counteract the effect of magnon dissipation and to find reliable knobs to sustain the magnon current for long-distance transport.
It has been proposed that magnon Bose-Einstein condensates \cite{Demo2006,Bozhko2019}, spin superfluids \cite{Takei2014,Wei2018}, spin Hall effect \cite{Gladii2016}, thermal spin torques \cite{Padron2011}, topological edge mode generation \cite{Malz2019}, and non-Hermitian coupling with cavity photons \cite{WangPRL2019} can be used to enhance magnon currents.

In this Letter, we show that the magnon spin current can be significantly amplified at an interface between a  magnet that is not driven externally and a magnet into which angular momentum is injected using spin-orbit torque (SOT) \cite{Miron2011,Liu2012,Garello2013,Manchon2015,Manchon2019}.
By designing the balance of this external driving with intrinsic dissipation, both magnons (positive-energy excitations) and antimagnons (negative-energy excitations) are dynamically stabilized. This results in enhanced reflection of magnons from the interface with the driven-dissipative magnet. The enhanced reflection is accompanied by a transmitted antimagnon current.
This suggests a method to amplify magnon spin currents that is relatively straightforward to implement, which may be generalized to both ferromagnetic and antiferromagnetic materials, different types of driving, and to both metals and insulators. Below, we explicitly illustrate the basic physics for a magnetic heterostructure involving yttrium-iron-garnet (YIG) and platinum.

Our result can be interpreted as a realization of the bosonic Klein paradox, which refers to the counterintuitive reflection or transmission of relativistic particles from a potential barrier \cite{Klein1929,Dombey1999,Alex1981,Nikishoiv1970,Gavrilov2016,Brito2021}, and is a natural consequence of relativistic quantum theory.
The experimental test of this paradox using fundamental particles is nearly impossible  because of the extremely high energy barrier that needs to be overcome \cite{bjork}.  While its solid-state realization for fermionic particles in 2D materials with gapless excitations was recently reported \cite{Kat2006,Stander2009,Chris2016}, the study of the Klein paradox for bosonic quasiparticles remains an outstanding challenge because the presence of bosonic antiparticles in a solid-state system usually signals instabilities. In our implementation, these instabilities are prevented by the external driving via SOT.
Hence, in addition to the application-motivated magnon amplification that is discussed above, our results launch driven magnetic systems as a suitable solid-state platform to experimentally study the relativistic physics of bosonic particles.

\begin{figure}
	\centering
	\includegraphics[width=0.48\textwidth]{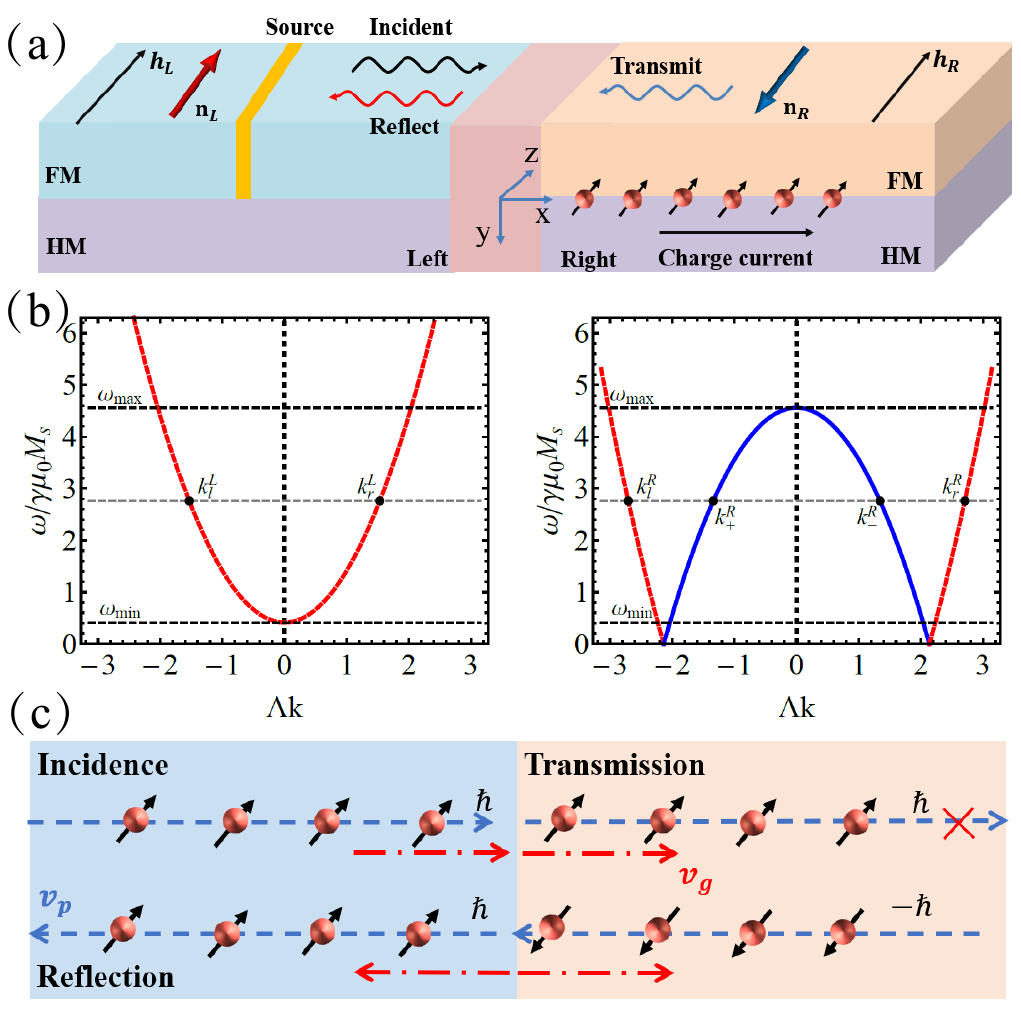}\\
	\caption{(a) Schematic of the driven-dissipative magnetic system containing two exchange-coupled magnetic films. (b) Magnon dispersion for the left and right ferromagnets respectively. The red curves give the positive energy excitations while the blue curve corresponds to the negative energy excitations. In other words, negative energy excitations exist for the wavenumbers $ -2.1\lesssim\Lambda k\lesssim2.1 $ between the zeros of the dispersion in the right magnet. (c) Physical picture of the anomalous magnon reflection. Blue and red arrows represent the directions of phase and group velocity, respectively.}\label{fig1}
\end{figure}

{\it Physical model.}---
We consider two exchange coupled ferromagnetic (FM) insulating thin films adjacent to a heavy-metal layer (HM) subject to an in-plane external magnetic field in the $ z $ direction, as shown in Fig. \ref{fig1}(a).
The magnetization of the right FM aligns antiparallel to the external field.
This situation is energetically unstable but dynamically stable due to the presence of an electrical current in the HM layer which exerts a SOT on the magnetization dynamics.
This can be understood as follows: the spin current produced by the electric current through the spin Hall effect in the HM will keep injecting angular momentum to the FM layer to counteract the damping of magnetization and to prevent the magnetization to align with the external field, thereby yielding a region with dynamically stable antimagnons.
In general, the dynamics of the magnetization $\mathbf{n}_\nu=\mathbf{M}_\nu/M_s $ is well described by the Landau-Lifschitz-Gilbert (LLG) equation with SOT~\cite{Manchon2019}, i.e.,
\begin{equation}\label{LLG}
	\frac{\partial \mathbf{n}_\nu}{\partial t}=-\gamma \mathbf{n}_\nu \times \mathbf{h}_{\mathrm{eff},\nu} + \alpha \mathbf{n}_\nu \times \frac{\partial \mathbf{n}_\nu}{\partial t} + J_\nu \mathbf{n}_\nu \times \hat{z} \times \mathbf{n}_\nu,
\end{equation}
where $ \nu=L,R $ labels the (L)eft and (R)ight magnet, and, $\gamma$ is the gyromagnetic ratio, $\alpha$ is the Gilbert damping and $J_\nu $ characterizes the strength of SOTs generated by the spin current which depends on the current flowing in the HM layer, the spin Hall angle of the HM and the properties of the interface.
The LLG equation describes damped precession around the effective magnetic field $\mathbf{h}_{\mathrm{eff},\nu}=-\delta E_\nu/(M_s\delta\mathbf{n}_\nu) $, with $M_s$ being saturation magnetization.
Here, we consider the magnetic energy functional $ E_\nu[\mathbf{n}_\nu] $ in the left and right magnet to be of the form
\begin{equation}\label{eq:effective_hamiltoniaan}
	\begin{aligned}
		E_{\nu}=&\int dV \bigg\{	
		A(\nabla_i\mathbf{n}_\nu)^2
		-\mu_0H_{\mathrm{e} ,\nu}M_s n_{z,\nu}+\frac{1}{2}Kn_{y,\nu}^2
		\bigg\},
	\end{aligned}
\end{equation}
with $ A $ the exchange stiffness,
$ H_{\mathrm{e},\nu} $ is the external magnetic field strength, $\mu_0 $ is the vacuum permeability and $ K=\mu_0M_s^2 $ the effective shape anisotropy caused by the dipolar interaction.

Spin waves or magnons are introduced as linear dynamical fluctuations on top of the equilibrium magnetization
$\mathbf{n}_{0,\nu}$ with $\mathbf{n}_{0,L}=e_z$ and $\mathbf{n}_{0,R}=-e_z$.
By introducing the complex field
\begin{equation}\label{eq:complex_field_Psi}
	\Psi_{\nu}=(1/\sqrt{2})\left(\hat{x}-\mathrm{i}\hat{y}\right)\cdot\mathbf{n}_{\nu},
\end{equation}
with
$ \mathbf{n}_\nu\simeq\hat{x}\,\sqrt{2}\mathrm{Re}[\Psi_\nu]+\hat{y}\,\sqrt{2}\mathrm{Im}[\Psi_\nu]
+ \mathbf{n}_{0,\nu}$,
the linearized LLG equation \cite{Kittel1947, SparksPR1970,Kalinikos1986} accordingly becomes
\begin{equation}\label{eq:linearised_LLG_equation}
	\begin{aligned}
		&\mathrm{i}(1\pm\mathrm{i}\alpha)\partial_t\Psi_{\nu}/\gamma\mu_0M_s
		\\&
		=
		\pm
		\left(\Delta\pm h_{\nu}-\Lambda^2\nabla^2+\mathrm{i}I_\nu\right)\Psi_{\nu}
		\pm\Delta\Psi_\nu^*,
	\end{aligned}
\end{equation}
with $ \Lambda=\sqrt{2A/\mu_0M_s^2} $ the exchange length,
$ h_{\nu}=H_{\mathrm{e},\nu}/M_s $ the dimensionless external magnetic field,
$ \Delta=K/2\mu_0M_s^2=1/2 $ the dimensionless anisotropy constant
and $ I_\nu=J_\nu/\gamma\mu_0M_s $ the dimensionless SOT. The $ \pm $ sign in the above equation comes from linearizing around the $ \pm e_z $ direction in the left and right ferromagnets respectively.

The solutions of the linearized LLG equation (\ref{eq:linearised_LLG_equation}) may be given in the form of Bogoliubov modes
\begin{equation}\label{eq:bogoliubov_ansatz}
	\Psi_{\nu}(\mathbf{x},t)=u_{\nu}(\mathbf{x})e^{-\mathrm{i}\lambda t}-v^*_{\nu}(\mathbf{x})e^{\mathrm{i}\lambda^*t}.
\end{equation}
The Fourier transform of $ u $ and $ v $ satisfy, up to first order in $ \alpha $ and $ I_\nu $, the following dispersion relation
\begin{equation}\label{eq:spinwave_dispersion}
	\omega_{\mathbf{k},\nu}
	\simeq
	\omega_{\mathbf{k},\nu}^0
	-\mathrm{i}(\alpha\omega_{\mathbf{k},\nu}^0+I_\nu),
\end{equation}
with $ \omega=\lambda/\gamma\mu_0M_s $ the dimensionless frequency and $(\omega_{\mathbf{k},\nu}^0)^2=(h_{\nu}\pm\Delta\pm\Lambda^2k^2)^2 - \Delta^2$.
In~\cref{fig1}(b) we plot the real part of the dispersion relation for both the left and right ferromagnets with $ h_L=0.5 $ and $ h_R=4.5$.

The energy functional in~\cref{eq:effective_hamiltoniaan} for solutions of~\cref{eq:linearised_LLG_equation} gives $ E_\nu=\int d\mathbf{k}\|\Psi_{\nu}\|_\mathbf{k}\cdot\omega_{\mathbf{k},\nu}^0 $, up to second order in $ u_\nu $ and $ v_\nu $, with $ \|\Psi_\nu\|_\mathbf{k}=\pm(|u_{\mathbf{k},\nu}|^2-|v_{\mathbf{k},\nu}|^2) $
~\cite{Lundh2006}.
Hence, energetic instabilities exist if the magnon excitation energy is negative, i.e.
$
\|\Psi_\nu\|_\mathbf{k}\cdot
\omega_{\mathbf{k},\nu}^0<0.
$
We thus note that the magnons on the right FM are energetically unstable for a range of wavenumbers, see~\cref{fig1}.
Physically, this means that the internal energy of the right ferromagnet can be lowered by small spin fluctuations (antimagnons).
In driven magnetic systems, energetic and dynamical stability do not necessarily coincide. For the system to be dynamically stable we need
$
\mathrm{Re}[(\mathrm{i}+\alpha)\|\Psi_\nu\|_\mathbf{k}\cdot\omega_{\mathbf{k},\nu}^0]+ I_\nu>0,
$
for all wavenumbers $ \mathbf{k}$, because then small-amplitude fluctuations die out. This identity imposes that the magnons on the right side are dynamically stable if $I_R\gtrsim\max\left[\alpha (h_R-\Delta),\Delta\right]$.

Additionally, the left and right thin films are exchange coupled \cite{noteinterface}, which results in effective boundary conditions for the magnetization.
In terms of the Bogoliubov ansatz~(\ref{eq:bogoliubov_ansatz}),
the four boundary conditions follow from varying the energy functional in~\cref{eq:effective_hamiltoniaan} after including the boundary term $E_{bn}=-J_c \mathbf{n}_L(0) \cdot \mathbf{n}_R(0)$ at the interface ($ x=0 $).
This gives
\begin{subequations}\label{eq:boundary-conditions-Bogoliubov-ansatz}
	\begin{align}
		\Lambda \partial_x\varphi_L-\Lambda_c(\varphi_R+ \varphi_L)=&0,\\
		\Lambda \partial_x\varphi_R+\Lambda_c(\varphi_L+ \varphi_R)=&0,
	\end{align}
\end{subequations}
with  $\varphi=u,v$ and $ \Lambda_c=J_c/\Lambda\gamma\mu_0M_s $. To analytically solve the scattering problem, we shall first focus on the isotropic case ($\Delta=0$) and further show the essential physics still holds for elliptical magnons ($\Delta\neq 0$).

{\it Scattering formalism.}---
Due to doubling of the modes we consider $ \omega>0 $ without loss of generality.
Solutions of~\cref{eq:linearised_LLG_equation} without dissipative terms have the form
\begin{equation}\label{eq:plain-wave-ansatz-bulk}
	\begin{pmatrix}
		u(x)\\v(x)
	\end{pmatrix}
	=
	\begin{pmatrix}
		u_k\\v_k
	\end{pmatrix}
	e^{ikx}.
\end{equation}
At a given $ \omega >\omega_{\mathrm{min}}\equiv h_L$, we find four different wavenumbers $ k $. In the left region with only positive energy excitations, we find two real $ k_r^L,k_l^L $ and two complex $ k_+^L,k_-^L $ wavenumbers.
 These wavenumbers are, according to~\cref{eq:spinwave_dispersion}, given by
\begin{equation}\label{eq:wavenumbers-left-magnet}
		\Lambda k^L_{r/l}
		=
		\pm\sqrt{\omega-h_L},
		\quad
		\Lambda k^L_{\pm}
		=
		\pm\mathrm{i}\sqrt{\omega+h_L}.
\end{equation}
The complex modes are either blowing up or are damped, where only the damped mode is physically allowed \cite{VerbaPRB2020}.
However, in the right magnet -- magnetized against the external magnetic field -- there are four real wavenumbers if $ \omega<\omega_{\mathrm{max}}\equiv h_R $, which are explicitly given by
\begin{equation}\label{eq:wavenumbers-right-magnet}
		\Lambda	k^R_{r/l}
		= \pm\sqrt{\omega+h_R},
		\quad
		\Lambda	k^R_{\pm}
		=\pm\mathrm{sgn}(\omega-h_R)\sqrt{h_R-\omega}.
\end{equation}
The $ k_r $ and $ k_l $ modes correspond to positive energy modes (magnons) with positive and negative group velocity respectively.
Furthermore, $ k_+ $ and $ k_- $ correspond respectively to additional right- and left moving modes carrying negative energy (antimagnons).
We included $ \mathrm{sgn}(\omega-h_R) $ in the expression of $ k_\pm $ here, such that $ k_+ $ corresponds both to the right moving negative energy mode and the exponentially damped mode when $ \omega>\omega_\mathrm{max}\equiv h_R $.

We now construct the scattering solutions satisfying the boundary conditions in~\cref{eq:boundary-conditions-Bogoliubov-ansatz}.
The general solution for bulk modes at frequency $ \omega $ are given by
\begin{subequations}\label{eq:bogoliubov-modes}
	\begin{align}
		u_\nu(x)=&A^\nu_{r}u_{k_r,\nu}e^{ik_rx}+A^\nu_{l}u_{k_l,\nu}e^{ik_lx}
		\\\nonumber
		+&A^\nu_{+}u_{k_+,\nu}e^{ik_+x}+A^\nu_{-}u_{k_-,\nu}e^{ik_-x},\\
		v_\nu(x)=&A^\nu_{r}v_{k_r,\nu}e^{ik_rx}+A^\nu_{l}v_{k_l,\nu}e^{ik_lx}
		\\\nonumber
		+&A^\nu_{+}v_{k_+,\nu}e^{ik_+x}+A^\nu_{-}v_{k_-,\nu}e^{ik_-x},
	\end{align}
\end{subequations}
with $ A^{\nu}_j $ are amplitudes of the scattering modes in~\cref{eq:wavenumbers-left-magnet,eq:wavenumbers-right-magnet} and $ \begin{pmatrix} u_{k,\nu},&v_{k,\nu} \end{pmatrix} $ solutions to~\cref{eq:linearised_LLG_equation} with ansatz~(\ref{eq:plain-wave-ansatz-bulk}) and normalization condition
$ ||u_{\nu,k}|^2-|v_{\nu,k}|^2|=1 $.
By disregarding spatially growing modes, the boundary conditions in~\cref{eq:boundary-conditions-Bogoliubov-ansatz} for incident magnons from the left imply
\begin{equation}\label{eq:scattering-equation-incoming-left}
	\mathbf{M}
	\begin{pmatrix}
		1\\A^L_l\\0\\A_-^L
	\end{pmatrix}
	=
	\begin{pmatrix}
		A_v^R\\0\\A_+^R\\0
	\end{pmatrix}.
\end{equation}
Here, the matrix $ \mathbf{M} $ is defined by the boundary conditions given in~\cref{eq:boundary-conditions-Bogoliubov-ansatz}
and is given by
\begin{align}
	&\begin{aligned}
		\mathbf{M}
		=
		\frac{1}{\lambda^R_{v,l}-\lambda^R_{v,r}}
		\begin{pmatrix}
			0&1\\
			0&0
		\end{pmatrix}
		&\bigotimes
		\begin{pmatrix}
			\lambda^R_{v,l}\lambda^L_{v,+}-1&\lambda^R_{v,l}\lambda^L_{v,-}-1\\
			1-\lambda^R_{v,r}\lambda^L_{v,+}&1-\lambda^R_{v,r}\lambda^L_{v,-}
		\end{pmatrix}
		\\
		+\frac{1}{\lambda^R_{u,-}-\lambda^R_{u,+}}
		\begin{pmatrix}
			0&0\\
			1&0
		\end{pmatrix}
		&\bigotimes
		\begin{pmatrix}
			\lambda^R_{u,-}\lambda^L_{u,r}-1&\lambda^R_{u,-}\lambda^L_{u,l}-1\\
			1-\lambda^R_{u,+}\lambda^L_{u,r}&1-\lambda^R_{u,+}\lambda^L_{u,l}
		\end{pmatrix},
	\end{aligned}
\end{align}
with
\begin{subequations}
	\begin{align}
		\lambda_{\varphi,j}^{L}=&
		+\mathrm{i}\Lambda_c^{-1}\Lambda k_j-1,\\
		\lambda_{\varphi,j}^{R}=&
		-\mathrm{i}\Lambda_c^{-1}\Lambda k_j-1.
	\end{align}
\end{subequations}
By solving (\ref{eq:scattering-equation-incoming-left}), we derive the reflection amplitudes as,
\begin{equation}\label{eq:reflection-coefficient}
		A^L_l=-\frac
		{1-\lambda^R_{u,+}\lambda^L_{u,r}}
		{1-\lambda^R_{u,+}\lambda^L_{u,l}},
		\quad
		A^L_-=0.
\end{equation}

We want to find the ratio between the incoming and reflected magnon spin current.
Here, we define the spin current as the spatial current following from the conservation of the norm -- without dissipative terms -- $ \|\Psi\|=|u|^2-|v|^2 $, i.e., $ -\mathrm{i}\partial_t\|\Psi\|\mp\Lambda\mathrm{i}\partial_xJ_s=0 $.
Using the equations of motion (\ref{eq:linearised_LLG_equation}), we find that the spin current
is given by $\mathrm{i}J_s\Lambda=u\partial_xu^*-u^*\partial_xu+v\partial_x v^*-v^*\partial_x v$.
Far from the interface, the Fourier transform of the spin current is dominated -- for wave packets -- by
\begin{equation}
	\begin{aligned}\label{eq:spin-current-definition}
		J_s/\Lambda=\sum_{k_j}
		|A_j|^2k_j
		\left(
		|u_{k_j}|^2
		+|v_{k_j}|^2
		\right).
	\end{aligned}
\end{equation}
Finally, we derive the reflection coefficient as the ratio of reflected and incident spin currents, by combining \cref{eq:wavenumbers-left-magnet,eq:reflection-coefficient,eq:spin-current-definition},
\begin{equation}\label{eq:reflected-spin-current}
	R^2\equiv-J_s^R/J_s^I=|A^L_l|^2.
\end{equation}

We now distinguish between the cases
$
\omega>
h_R
$
and
$
h_L
<\omega<
h_R
$.
(i) If $ \omega>h_R$ then $ |A_l^L|^2=1 $, hence we have perfect reflection in this instance.
(ii) If $ h_L<\omega<h_R $, then the reflection coefficient
	\begin{align}\label{eq:reflection-spin-current}
		&R^2
		=
		\\\nonumber
		&\frac
		{h_R-h_L+\Lambda_c^{-2}(\omega-h_L)(h_R-\omega)+2\sqrt{\omega-h_L}\sqrt{h_R-\omega}}
		{h_R-h_L+\Lambda_c^{-2}(\omega-h_L)(h_R-\omega)-2\sqrt{\omega-h_L}\sqrt{h_R-\omega}}.
	\end{align}
For $ \Lambda_c>\sqrt{h_R-h_L}/2$, the above expression is maximal at $ \omega=(h_L+h_R)/2 $ with the maximal reflection
\begin{equation}
	R^2_\mathrm{max}
	=
	1+8\Lambda_c^2/(h_R-h_L).
\end{equation}
On the other hand, if $\Lambda_c<\sqrt{h_R-h_L}/2$, the expression (\ref{eq:reflection-spin-current}) is maximal for $ \omega=(h_R+h_L)/2\pm\sqrt{(h_R+h_L)(h_R+h_L-4\Lambda_c^2)/4} $ with the maximal reflection
\begin{equation}\label{key}
	R^2_\mathrm{max}
	=
	\frac
	{\sqrt{h_R-h_L}+|\Lambda_c|}
	{\sqrt{h_R-h_L}-|\Lambda_c|}.
\end{equation}
For both cases, we find the maximum reflection $R_\mathrm{max}>1$, which gives rise to spin wave amplification and is the magnonic analogue of the Klein paradox.

\begin{figure}
  \centering
  \includegraphics[width=0.48\textwidth]{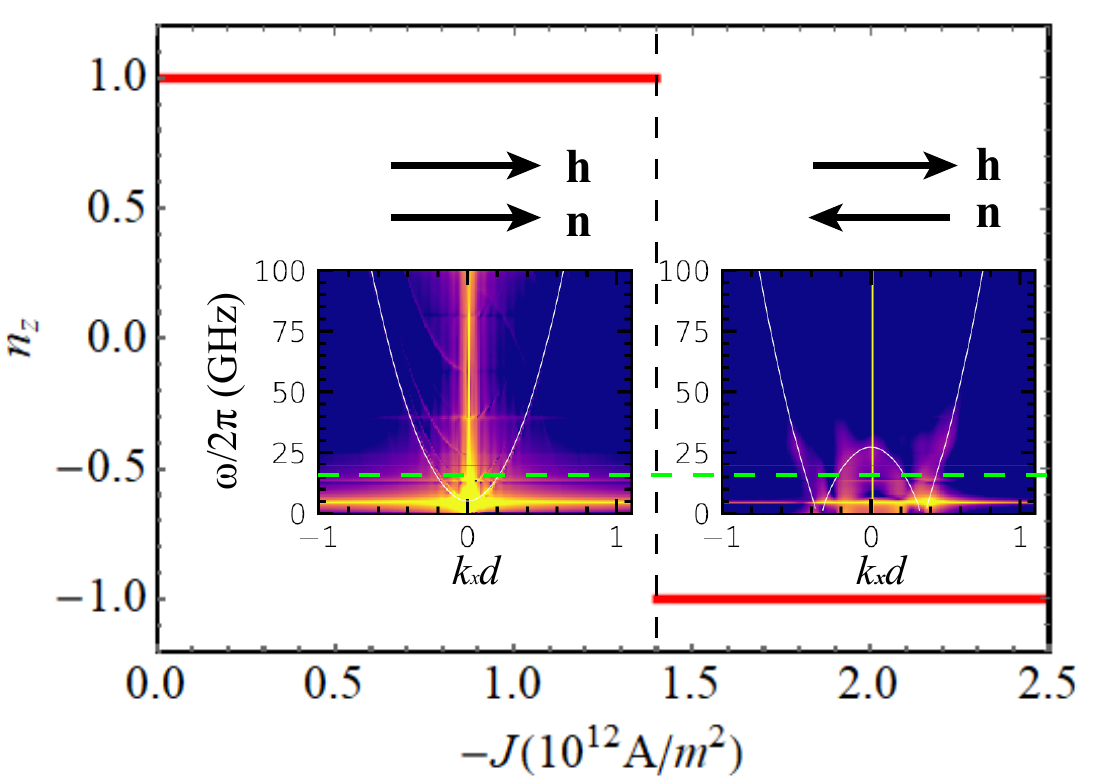}\\
  \caption{Steady state $n_z$ of the right magnet as a function of current density ($J$) obtained by simulations. The insets shows the simulated magnon spectrum for $J=0$ and $-1.5 \times 10^{-12}~ \mathrm{A/m^2}$, respectively. The white lines are the analytical dispersions. The horizontal dashed line is the driving frequency of the microwave ($\omega_0/2\pi = 20$ GHz) to initiate magnon scattering.}\label{fig2}
\end{figure}

The physical picture of this anomalous reflection is illustrated in Fig. \ref{fig1}(c). Magnons with angular momentum (AM) $\hbar$ are excited in the left domain and incident at the interface. The overlap of the magnon band in the left magnet with the antimagnon band in the right magnet, produced by the unequal external fields, guarantees that the magnons can propagate into the right film. According to AM conservation, a magnon current with AM $\hbar$ propagating forward is expected to be produced in the right domain. However, magnons with AM $\hbar$ are forbidden due to the antiparallel orientation of magnetization with respect to external field. To conserve AM, antimagnon currents with AM $-\hbar$ propagating backward are generated and thus enlarge the reflected current. Throughout the scattering process, the group velocity of transmitted antimagnons ($v_g$) is always positive to guarantee the forward flow of energy. As a comparison, in the original Klein paradox, an electrostatic potential lifts the positron band in the right region and makes it overlap with the electron band in the incident region, while in the present case an inhomogeneous field lifts the antimagnon band and makes it overlap with the magnon band \cite{notesm}.

{\it Numerical verification.}---To verify the analytical predictions and to account for effects of dissipation and non-linearities, we perform micromagnetic simulations \cite{notesm,Wang2014,mumax} on two exchange-coupled ferromagnetic thin films as shown in Fig. \ref{fig1}(a). Here the inter-domain coupling $A_i$ is a tunable coefficient and is related to $J_c$ in the theory as $J_c=-2A_i/d$.
By applying a global driving microwave $\mathbf{h}(t)=h_0 \mathrm{sinc} (\omega_c t)\hat{x}$ with $\omega_c/2\pi=100$ GHz and $h_0=50$ mT, we first quantify the response of the magnetic system and identify two regimes as shown in Fig. \ref{fig2}. (i) When the current density $J>- 1.4 \times 10^{12} ~\mathrm{A/m^2}$ , the antiparallel state of the right domain ($\mathbf{n}_R \parallel -\mathbf{h}_R$) is dynamically unstable and the magnetization switches to the parallel state spontaneously ($\mathbf{n}_R \parallel\mathbf{h}_R$). The magnon spectrum of the right magnet for the steady parallel state is a normal parabola (left inset of Fig. \ref{fig2}) \cite{notestanding}. (ii) When $J<- 1.4 \times 10^{12} ~\mathrm{A/m^2}$, the antiparallel state becomes dynamically stable and the antimagnon states in the negative energy branch are excited, and a sombrero-like spectrum is identified (right inset of Fig. \ref{fig2}), consistent with the theory.

\begin{figure}
  \centering
  \includegraphics[width=0.46\textwidth]{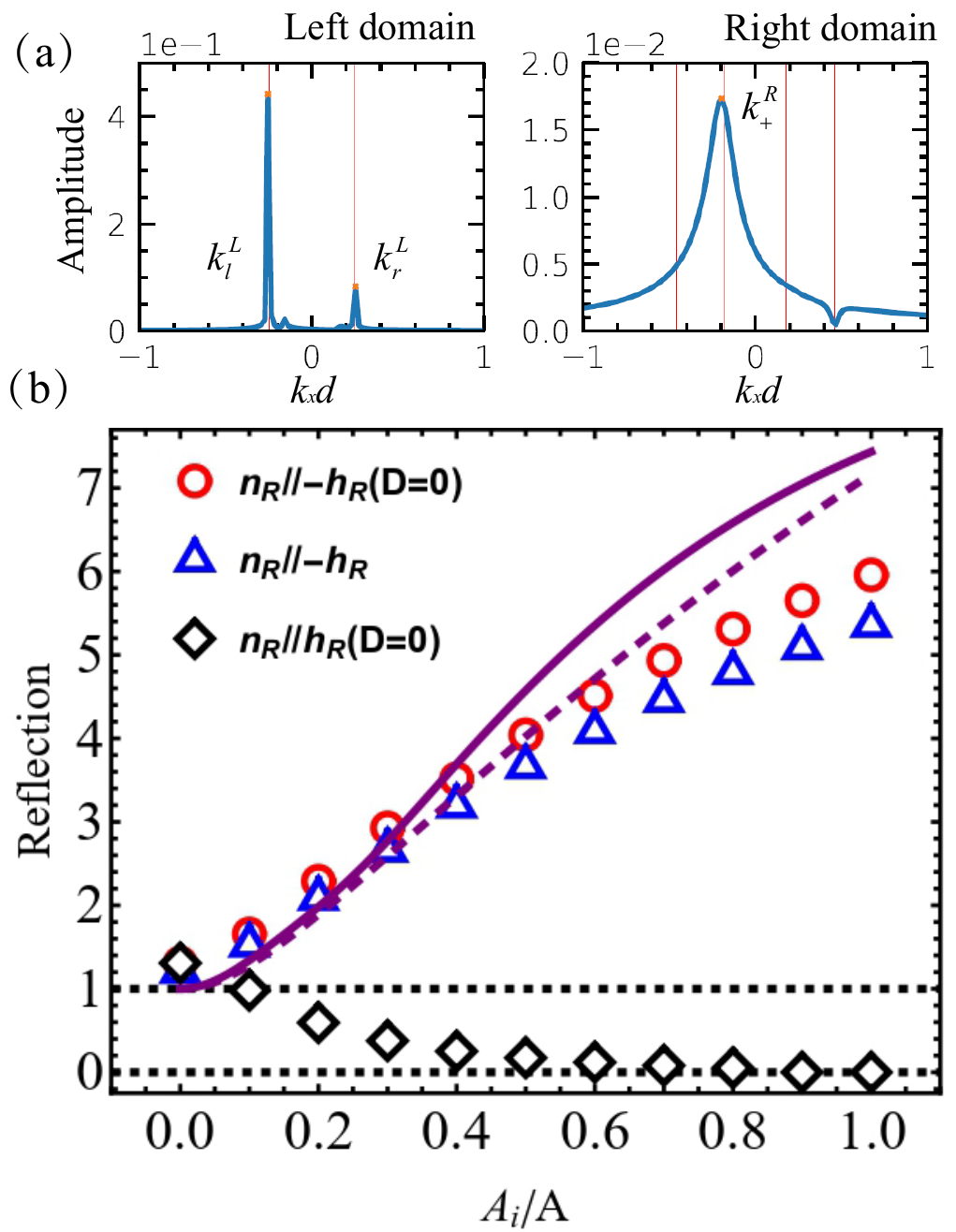}\\
  \caption{(a) Scattering of magnons at the interface of left and right domains. $h_0=10 ~\mathrm{mT}$, $A_i/A=1$. The red vertical lines represent the theoretical predictions of magnon wavevectors. (b) Reflection of magnons as a function of inter-domain exchange couplings. The DMI strength is $D=0$ (red circles) and $D=0.1~\mathrm{mJ/m^2}$ \cite{WangPRL020} (blue triangles). The purple dashed line is prediction by Eq. (\ref{eq:reflection-spin-current}), and the solid line is the prediction with shape anisotropy.}\label{fig3}
\end{figure}

To study the magnon scattering off the interface between left and right magnets, a microwave source $\mathbf{h}(t)=h_0\sin (\omega_0 t)\hat{x}$ is applied at the left domain at $x=-d_{si}$, with $d_{si}=800~\mathrm{nm}$. The excited magnons propagate in the $+\hat{x}$ direction and scatter at the interface ($x=0$).
By making a Fourier transform of $n_x(x,y,t)$ in the propagating direction ($\hat{x}$) \cite{notefft}, we derive the response of the system in momentum space as shown in Fig. \ref{fig3}(a). An antimagnon state with $k_x<0$ is clearly identified in the right domain while the reflection coefficient is larger than one.
This demonstrates the enhancement of magnon spin current via an analogue of the Klein paradox. In the absence of injection, the antimagnon current is barely excited. A detailed analysis of the evolution of incident, transmitted and reflected magnon current further verifies their correlation \cite{notesm}.

Figure \ref{fig3}(b) shows that the reflected coefficient, defined as the peak-height ratio of the reflection magnons and incident magnons, increases with inter-domain exchange coupling.
For comparison, we also simulate the magnon scattering in the parallel configuration ($\mathbf{n}_R \parallel \mathbf{h}_R$) and find that the reflection keeps decreasing to zero with increasing the coupling between the magnetic films (black diamonds). As expected, no antimagnon state is excited in this case.
We find a good agreement between the analytical prediction in~Eq.~(\ref{eq:reflection-spin-current}) and the micromagnetic simulations for small couplings. For large exchange couplings, however, we see quantitative differences, which are not explained by including dipolar anisotropy -- see purple solid line in~\cref{fig3}(b).
We expect the quantitative difference at large couplings to stem from non-linear effects, which are not treated in the analytical formalism. The reflection amplitudes become increasingly large at increasing couplings, thereby making non-linear effects important.

{\it Discussions and conclusions.}--- In conclusion, we have analytically shown and numerically confirmed that the magnon spin current can be amplified through the realization of the bosonic Klein paradox in a driven-dissipative magnetic system.  The Dzyaloshinskii-Moriya interaction (DMI) caused by the interfacial symmetry breaking in the hybrid system does not change the results significantly, as shown in Fig. \ref{fig3}(b). In our proposal, we dynamically stabilize the antimagnons by the SOT. The essential physics is applicable to a wide class of materials and driving knobs which are able to maintain the magnetization against the external field. For example, electric currents through spin-transfer torque \cite{Wegrowe2007}, optical waves through magneto-optical interaction \cite{Cao2020} and other effective techniques capable of producing a positive damping of the magnons. Our proposal therefore can be realized in ferromagnetic insulators as well as metals. Experimentally, the magnons may be detected by optical, inductive and even electric techniques \cite{Seb2015, Vlaminck2010,Corn2015}.

\begin{acknowledgments}
 H.Y.Y acknowledges the European Union's Horizon 2020 research and innovation programme under Marie Sk{\l}odowska-Curie Grant Agreement SPINCAT No. 101018193. R.A.D is member of the D-ITP consortium, a program of the Netherlands Organisation for Scientific Research (NWO) that is funded by the Dutch Ministry of Education, Culture and Science (OCW). R.A.D. acknowledges the funding from the European Research Council (ERC) under the European Union's Horizon 2020 research and innovation programme (Grant No. 725509).
 This work is part of the Fluid Spintronics research programme with project
 number 182.069, which is financed by the Dutch
 Research Council (NWO).
\end{acknowledgments}

{}
\pagebreak
\pagestyle{empty}
\setlength{\textheight}{\paperheight}
\setlength{\textwidth}{\paperwidth}
\setlength{\headheight}{1.75cm}
\setlength{\voffset}{-2cm}
\setlength{\hoffset}{-2cm}
\onecolumngrid
\centering
\includegraphics[page=1]{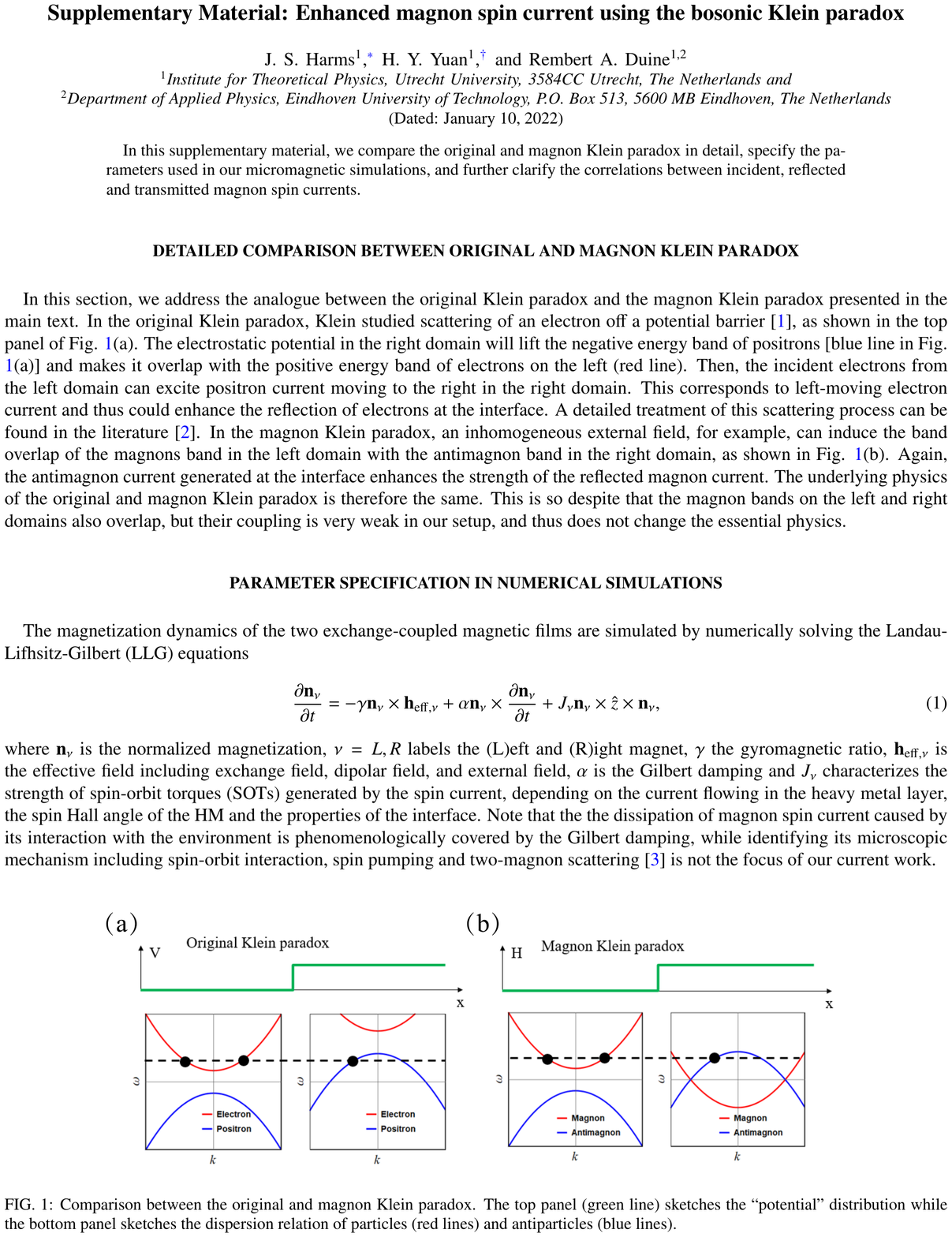}
\includegraphics[page=2]{Supplement}
\includegraphics[page=3]{Supplement}

\begin{thebibliography}{99}
\bibitem{noteinterface} Here the interface of left and right domains may be a normal-metal layer that generates the interlayer exchange interaction between left and right domain, an irradiated region where magnetic parameters can be effesitively designed or other regions that could induce an effective exchange coupling between left and right domains.
\bibitem{notestanding} Here the lineshape of dispersion is consistent with the theoretical predictions (white lines), while we also notice a series of standing wave modes in the transverse direction ($k_y w=n \pi$, $n=0,1,2$,...). For a real magnetic film with a sufficiently larger width, the lower modes of the standing waves will merge together. This will not have a significant influence on our results.
\bibitem{notefft} Here to eliminate the influence of magnons on the regions $x<-d_i$, which have the same wavevector and propgating direction as the reflecting magnons, the FFT is only performed at the region between the microwave source and the interface while same sized region is taken in the right domain.
\end{thebibliography}
\end{document}